\documentclass{aastex}
\usepackage{spr-astr-addons}

\begin{document}

\title{Discovery of deep eclipses in the cataclysmic variable ZTF17aaaeefu 
(2MASS~J00594349+6454419)}
\shorttitle{Eclipses in ZTF17aaaeefu}
\shortauthors{Kozhevnikov}

\author{V. P. Kozhevnikov}
\affil{Astronomical Observatory, Ural Federal University, Lenin Av. 51, Ekaterinburg 
620083, Russia e-mail: valery.kozhevnikov@urfu.ru}

\begin{abstract} 

I performed photometric observations of the cataclysmic variable candidate ZTF17aaaeefu and discovered very deep eclipses. The observations were obtained over 8 nights covering 7 months. The eclipse profile is similar to the eclipse profiles observed in other cataclysmic variables. During the observations, ZTF17aaaeefu showed brightness changes resembling dwarf nova outbursts and quiescent states. When ZTF17aaaeefu was bright (16.0--16.2~mag), the average eclipse depth was $2.50\pm0.18$~mag. When ZTF17aaaeefu was faint (17.2--17.4~mag), the average eclipse depth was $1.45\pm0.06$~mag. However, these differences in the eclipse depth may be mainly caused by the contaminating effect of three faint stars around ZTF17aaaeefu.  In both cases, the average width of the prominent parts of the eclipses was the same and was about 30~min. Due to the large coverage of observations, I measured the orbital period with high precision, $P_{\rm orb}=0.188\,211\,55\pm0.000\,000\,14$~d. I derived the eclipse ephemeris, the validity time of which is 700 years in accordance with the precision of the orbital period. This ephemeris can be used for future studies of the orbital period changes. Because ZTF17aaaeefu has a long orbital period, this cataclysmic variable is of interest for determining the masses of its stellar components. In future radial velocity measurements, my precise eclipse ephemeris may be useful for determining the orbital phases.

\end{abstract}


\section{Introduction}

Cataclysmic variables (CVs) are interacting binary stars consisting of a white dwarf and a late-type companion that fills its Roche lobe and transfers material to the white dwarf. Depending on the presence or absence of large outbursts, CVs are divided into dwarf novae and nova-like variables. Classical and recurrent novae with the exception of symbiotic stars also belong to CVs. The magnetic field of a white dwarf can greatly affect the accretion process. Accretion occurs through an accretion disc in non-magnetic systems, whereas it occurs through a truncated accretion disc or through an accretion column in magnetic systems. Magnetic systems divided into polars, in which the white dwarf rotates synchronously with the orbit, and intermediate polars, which reveal coherent brightness oscillations due to the non-synchronous rotation of the white dwarf. All CVs exhibit non-periodic brightness changes that have time scales ranging from seconds to several tens of minutes. This phenomenon is called flickering. The accretion luminosity in all CVs is much higher than that of the stellar components. Therefore, their stellar components are difficult to detect. If the inclination of the orbit is high ($i > 60^\circ$), CVs show eclipses. Eclipsing CVs are important because eclipses allow us to reliably determine the orbital period and the inclination of the orbit, as well as to study the structure of the accretion disc or accretion column. Comprehensive reviews of CVs are given in \citet{ladous94}, \citet{warner95}, \citet{hellier01}. Some resent data about eclipsing CVs one can found in \citet{hardy17}.

\citet{szkody20} identified 218 previously unknown objects as strong candidate CVs based on the shape and colour of their light curves. Only a few of these candidates were brighter than 19~mag in low-brightness states, which allowed their photometry to be carried out using a relatively small telescope. One of them, ZTF17aaaeefu (hereafter ZTF17), due to its favourable coordinates RA=$00^h59^m43.49^s$ and Dec=$+64^\circ54'41.9''$ can be observed throughout the year at our observatory near Yekaterinburg. From Table 2 in \citeauthor{szkody20}, I learned that ZTF17 showed 6 outbursts during 250 days. Hence, this object may be a dwarf nova. I performed tentative photometric observations of ZTF17 in 2020 September 10 when this star was quite bright (16~mag). I immediately discovered a very deep eclipse (2.9~mag). Eclipses in ZTF17 were not previously known. To measure the eclipse period with high precision, I conducted photometric observations of ZTF17 over 8 nights covering 7 months. In this paper, I present the results of these observations.

\section{Observations}

I conducted photometric observations of ZTF17 at Kurovka observatory, Ural Federal University, which is located 80 km from Yekaterinburg. I used a 70-cm Cassegrain telescope, equipped with a three-channel pulse-counting photometer with photomultipliers. The design of the photometer is described in \citet{kozhevnikoviz}. The photometer allows me to measure the brightness of two stars and the sky background. This makes it possible to obtain high-quality photometric data in conditions of unstable atmospheric transparency and high and variable sky background. In addition, I use the CCD guiding system, which allows me to automatically maintain the centring of the two stars in the photometer diaphragms. This guiding system improves the accuracy of brightness measurements and, together with a PC-based data acquisition system, allows observations to be made automatically and nearly continuously. Short interruptions occur in accordance with a computer program to measure the sky background in all three channels simultaneously. This is necessary to determine the differences in the sky background that are caused by differences in diaphragm sizes and to eliminate the effect of faint stars in the sky background channel. This also allows me to exclude harmful effects caused by changing the sky background with wavelength and time.

Most photoelectric photometers enable to observe relatively bright stars that are visible to the eye. For our 70-cm telescope, the limit of visibility is 15~mag. A few years ago, however, I learned to observe much fainter stars. A faint star that is invisible to the eye can be placed in the diaphragm of the photometer using the coordinates of the faint star and the coordinates of the nearest reference star as well the step motors of the telescope. The comparison star should be visible to the eye. Although each star with known coordinates can be placed in the photometer diaphragm, it should be noted that the centring of stars fainter than 18~mag is difficult to verify, whereas the centring of brighter stars can be verified and corrected if necessary using the telescope's step motors and star counts. However, the comparison star and program star centred once maintain relative centring throughout the observation season. For faint stars, Poisson fluctuations of light fluxes (photon noise) are significant. Nevertheless, using the three-channel photometer and  our 70-cm telescope, I was able to discover and measure deep eclipses in two CVs, in which the deepest points in the eclipses reached 19.5 and 20~mag \citep{kozhevnikov18, kozhevnikov20}. This would be impossible without the sky background channel.

\begin{table}[b]
{\small 
\caption{Log of the observations}
\label{table1}
\begin{tabular}{@{}l c c }
\hline
\noalign{\smallskip}
Date (UT) & BJD$_{\rm TDB}$ start & Length (h)\\ 
                 & (-2,459,000)                  &                 \\
\noalign{\smallskip}
\hline
\noalign{\smallskip}
2020 Sep. 10   & 103.217470    &   4.4 \\
2020 Sep. 18   & 111.180977    &   3.1 \\
2020 Sep. 24   & 117.205864    &   6.8 \\
2020 Sep. 25   & 118.199595    &   6.0 \\
2021 Apr. 6      & 311.210530    &   4.3 \\
2021 Apr. 16    & 321.224439    &   5.3 \\
2021 Apr. 17    & 322.216395    &   5.7 \\
2021 Apr. 18    & 323.255059    &   4.2 \\
\noalign{\smallskip}
\hline
\end{tabular} }

\end{table}


The first night of observations in 2020 September 10 showed that ZTF17 was quite bright (16~mag), so it was easy to verify its centring in the photometer diaphragm using star counts. 
This night allowed me to discover a very deep eclipse. In 2020 September, the observations were performed during three other nights until I obtained two pairs of consecutive eclipses, which allowed me to roughly measure the eclipse period. The next observations were delayed until 2021 April to measure the eclipse period with high precision. 

The observations of ZTF17 were performed over 8 nights with a total duration of 39.8~h. The data were obtained in white light (approximately 3000--8000~\AA) with a time resolution of 16~s. For ZTF17 and the comparison star, I used 16-arcsec diaphragms. For the sky background, I used a 30-arcsec diaphragm. This large diaphragm reduces the photon noise caused by the sky background. The comparison star is USNO-A2.0 1500-01027361. It has $B=13.4$~mag and $B-R=1.2$~mag. The colour index of ZTF17 is similar to the colour index of this star.  According to the USNO-A2.0 catalogue, ZTF17 has $B=17.4$~mag and $B-R=1.0$~mag.  Similar colour indexes of ZTF17 and the comparison star reduce the effect of differential extinction. The differential magnitudes were determined taking into account the differences in light sensitivity between the cannels of the photometer.

The log of the observations is shown in Table~\ref{table1}. It shows BJD$_{\rm TDB}$, which is the  Barycentric Julian Date in the Barycentric Dynamical Time (TDB) standard. BJD$_{\rm TDB}$ is uniform and therefore preferred. I used the online calculator (http://astroutils.astronomy.ohio-state.edu/time/) \citep{eastman10} to calculate BJD$_{\rm TDB}$. In addition, I calculated BJD$_{\rm UTC}$ using the BARYCEN routine in the 'aitlib' IDL library of the University of T\"{u}bingen (http://astro.uni-tuebingen.de /software/idl/aitlib/) to verify BJD$_{\rm TDB}$. During my observations of ZTF17, the difference between BJD$_{\rm TDB}$ and BJD$_{\rm UTC}$ was constant. BJD$_{\rm TDB}$ exceeded BJD$_{\rm UTC}$ by  69~s (e.g., \citealt{eastman10}).

\section{Analysis and results}

\begin{figure}[t]
\includegraphics[width=84mm]{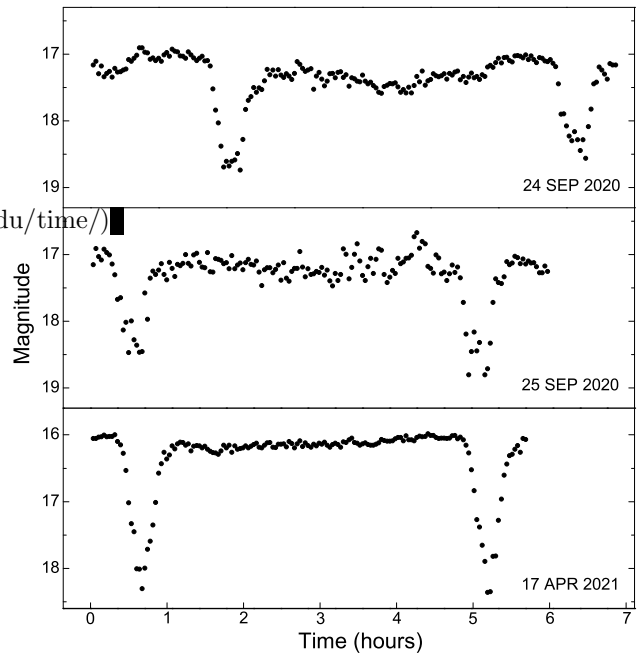}
\caption{Three longest light curves of ZTF17, which show consecutive eclipses. Note that the out-of-eclipse magnitude varies greatly from night to night} 
\label{figure1}
\end{figure}

All eight long differential light curves of ZTF17 obtained using the multichannel photometer (Table~\ref{table1}) clearly showed deep eclipses. Initially, I used a time resolution of 16~s. The photon noise of the differential light curves outside the eclipses (rms) was in the range 0.06--0.18~mag. It varied greatly depending on the brightness of ZTF17 and the sky background. However, at a time resolution of 16~s, the differential magnitudes in the innermost parts of the eclipses fluctuated greatly, or they could not be determined at all. So I averaged the counts over 128-s time intervals. Fig.~\ref{figure1} shows three longest light curves of ZTF17 obtained with a time resolution of 128~s, in which pairs of consecutive eclipses are obvious. The photon noise of these light curves outside the eclipses is 0.02--0.06~mag, the photon noise in the innermost parts of the eclipses is 0.2--0.4~mag, and the differential magnitudes of all eclipse points are determined. So I used a time resolution of 128~s to analyse the eclipses. 

The relative sky background for the innermost points of the eclipses was very high (96--98 \%). Therefore, these points may be shifted up or down due to inaccurate subtraction of the sky background. As seen in the upper, middle and lower frames of Fig.~\ref{figure1}, consecutive eclipses have comparable depths, which differ by 0.24, 0.26 and 0.10~mag, respectively (see Table~\ref{table2}).  Thus, the errors of the innermost points of the eclipses caused by inaccurate subtraction of the sky background do not exceed several tenths of a magnitude.

In Fig.~\ref{figure1} and in other suitable cases, for ease of use, I converted the differential magnitudes to the magnitudes by adding the $G$-magnitude of the comparison star (Gaia~EDR3~524277861729737216, $G$=12.22~mag) \citep{gaia16, gaia18}. The $G$-magnitudes measured using the space observatory {\it Gaia} seem to be the most appropriate because the spectral responses of the photomultipliers (S20 photocathodes) seem to be compatible with the $G$-band, and my white light magnitudes seem to be compatible with the $G$-magnitudes (see, e.g., \citealt{maiz18}). In addition, as mentioned, the comparison star has the colour index similar to the ZTF17 colour index. This reduces the effect of differences in the spectral bands.

\subsection{Determination of eclipse parameters}

As seen in Fig.~\ref{figure1}, the magnitude of ZTF 17 outside the eclipses shows large changes from night to night. The out-of-eclipse magnitudes on all the observation nights are shown in Table~\ref{table2}. They were determined by averaging significant portions of the out-of-eclipse light curves. Fig.~\ref{figure1} also shows that, depending on the brightness of ZTF 17, the eclipses have different depths. During the observations, I obtained 11 eclipses. I divided these eclipses into three groups. 5 eclipses were of high quality when the eclipses were more deeper and the eclipse points showed a small scatter (the out-of-eclipse magnitudes were 16.0--16.2~mag), 5 eclipses were of low quality when the eclipses were less deeper and the eclipse points showed a large scatter (the out-of-eclipse magnitudes were 17.2--17.4~mag), and 1 eclipse was of moderate quality (the out-of-eclipse magnitude was 16.9~mag).

The parameters of the observed eclipses are shown in Table~\ref{table2}. The mid-eclipse time was determined using a Gauss function fit to the eclipse (e.g., \citealt{groot98}), where the eclipse was cut off at 80--90 \% of the eclipse depth to eliminate the effect of asymmetric eclipse wings. In contrast, the depth of the eclipse was determined using a Gauss function fit to the entire eclipse and two adjacent parts of the out-of-eclipse light curve, the length of each of which was roughly the same as the length of the entire eclipse. As seen in Table~\ref{table2}, the errors determined from the fit depend on the brightness of ZTF 17 and the eclipse depth and, therefore,  on the eclipse quality. From Table~\ref{table2}, I determined that the average eclipse depth was $2.38\pm0.19$~mag for high- and moderate-quality eclipses and $1.45\pm0.06$~mag for low-quality eclipses.

\begin{table}[t]
\caption[ ]{Parameters of the observed eclipses}
\scriptsize
\label{table2}
\begin{tabular}{@{}l c l c}
\hline
\noalign{\smallskip}
Date (UT)  & Out-of-ecl.    & BJD$_{\rm TDB}$ mid-ecl.   &   Ecl. depth  \\
                  & magnitude    & (-2,459,000)                          &   (mag)       \\
\noalign{\smallskip}
\hline
\noalign{\smallskip}
2020 Sep. 10       & 15.99(2)     & 103.35356(18)    & 2.86(5)    \\
2020 Sep. 18       & 17.20(4)     & 111.25910(50)    & 1.35(9)    \\
2020 Sep. 24a     &  17.36(6)    & 117.28177(27)    &  1.62(8)   \\
2020 Sep. 24b     &  17.36(6)    & 117.47005(37)    & 1.38(6)    \\
2020 Sep. 25a     &  17.24(2)    & 118.22241(47)    & 1.32(9)    \\
2020 Sep. 25b     &  17.24(2)    & 118.41034(55)    & 1.58(11)  \\
2021 Apr. 6          &  16.19(1)    & 311.32775(33)    & 2.93(9)   \\
2021 Apr. 16        &  16.90(7)    & 321.30262(31)    & 1.76(6)   \\
2021 Apr. 17a      &  16.10(4)    & 322.24379(18)     & 2.06(5)   \\
2021 Apr. 17b      &  16.10(4)    & 322.43244(21)     & 2.15(5)   \\
2021 Apr. 18       &  15.99(8)     &  323.37294(13)    & 2.52(5)  \\
\noalign{\smallskip}
\hline
\end{tabular}
\end{table}

\subsection{Determination of the orbital period}

The three pairs of consecutive eclipses shown in Fig.~\ref{figure1} allow me to determine the approximate eclipse period, which is obviously equal to the orbital period. Although measuring the time interval between consecutive eclipses cannot provide high period precision, it is useful because such a measurement eliminates any aliasing problem that may arise as a result of complex analysis. I determined three periods from the precise mid-eclipse times of these three pairs of eclipses. They are shown in Table~\ref{table3}. The average orbital period is $0.18829\pm0.00021$~d.


\begin{table}[b]
{\scriptsize
\caption{Determination of the period from pairs of consecutive eclipses}
\label{table3}
\begin{tabular}{@{}l c c c}
\hline
\noalign{\smallskip}
Date (UT)  &\tiny BJD$_{\rm TDB}$ mid-ecl.1 &\tiny BJD$_{\rm TDB}$ mid-ecl.2  & Period (d)\\
                  &  (-2,459,000)                              &          (-2,459,000)                          &        \\
\noalign{\smallskip}  
\hline
\noalign{\smallskip}
2020 Sep. 24    & 117.28177(27)   & 117.47005(37)   & 0.18828(46)   \\ 
2020 Sep. 25    & 118.22241(47)   & 118.41034(55)   & 0.18793(72)   \\
2021 Apr. 17    & 322.24379(18)    & 322.43244(21)   & 0.18865(28)   \\
\noalign{\smallskip}
\hline
\noalign{\smallskip}
                                                             & & Average &                        \\
                                                             & & period    : & 0.18829(21)   \\
\noalign{\smallskip}
\hline
\end{tabular} }
\end{table}

\begin{figure}[t]
\includegraphics[width=84mm]{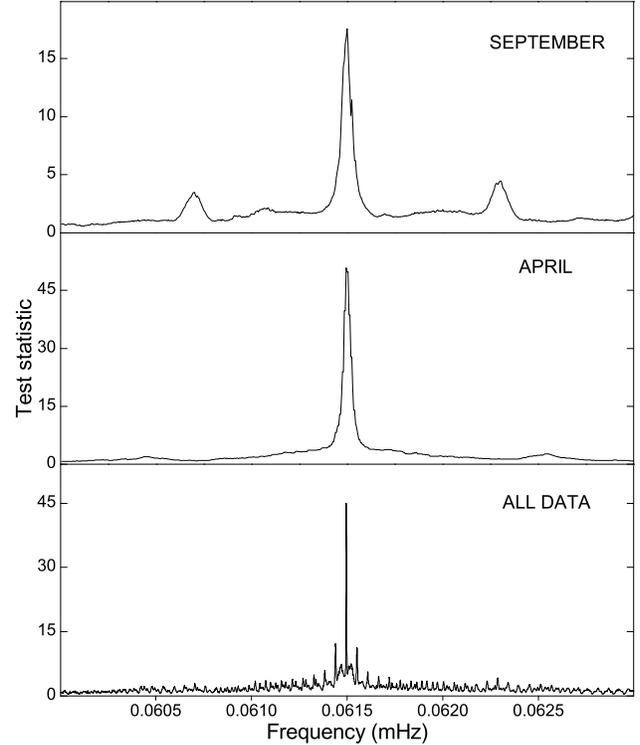}
\caption{Analysis of Variance spectra of ZTF17 calculated for the data obtained in 2020 September, for the data obtained in 2021 April and for all the data combined. Note that the main peak in the AoV spectrum of all the data is much narrower}
\label{figure2}
\end{figure}

For eclipsing variables, it is believed that fitting the ephemeris to the eclipse times gives the best precision of the orbital period. Obviously, if the observations cover a large time span, the ephemeris used for fitting should be sufficiently precise to avoid ambiguity in the cycle count. I used the Analysis of Variance (AoV) method \citep{schwarzenberg89} to determine the orbital period with sufficient precision. This method is advantageous compared to the Fourier transform for non-sinusoidal signals \citep{schwarzenberg98}. Usually, the AoV method is applied to data folded with trial periods. Instead, I assign trial frequencies with constant change steps, and then apply the AoV method. Thus, instead of an AoV periodogram, I obtain an AoV spectrum that covers a much larger time interval of variability. 

For the Fourier power spectrum, \citet{schwarzenberg91} showed that the $1\sigma$ confidence interval of the oscillation period is equal to the peak width at the $S-N$ level, where $S$ is the peak height and $N$ is the average noise power level. Unfortunately, a simple method for determining the period error for AoV spectra is unknown. Therefore, I used the method by \citeauthor{schwarzenberg91} also for AoV spectra as a tentative one.

\begin{table}[b]
\scriptsize
\caption{Determination of the period from AoV spectra}
\label{table4}
\begin{tabular}{@{}l c c c c }
\hline
\noalign{\smallskip}
Time span & S/N       & HWHM of       & Period (d) & deviation          \\
                  &              & the peak (d)   &                  & from all data     \\
\noalign{\smallskip}  
\hline
\noalign{\smallskip}
2020 Sep. &  18    &  0.00020    &  0.188224(24)    & $0.5\sigma$    \\ 
2021 Apr.  &  48    &  0.00014    &  0.188211(10)     & $0.1\sigma$    \\
All data      &  43   &  0.000010  &  0.1882118(8)    & --        \\
\noalign{\smallskip}
\hline
\end{tabular} 
\end{table}

I calculated the AoV spectra for three groups of ZTF17 data, namely for the data obtained in September, for the data obtained in April and for all the data combined. Previously, the out-of-eclipse magnitudes were subtracted from the light curves. For all the data combined, which cover 7 months, the trial frequency steps were 0.000\,000\,25 mHz. For the data obtained in September and April, the trial frequency steps were 20 times less. The AoV spectra are shown in Fig.~\ref{figure2}. As seen, the noise level in the upper frame of Fig.~\ref{figure2} is noticeably higher than in other frames. The reason is that the data obtained in September contained 5 low-quality eclipses and only 1 high-quality eclipse. In contrast, the data obtained in April contained 4 high-quality eclipses and only 1 moderate-quality eclipse. Nevertheless, all the AoV spectra show distinct main peaks coinciding in frequency, which are much higher than the noise level. In addition, all aliases that are noticeable due to their symmetric location relative to the main peaks have much lower heights than the heights of the main peaks. Thus, despite the large gap between the September and April data, the AoV spectrum of all the data combined shows no problems with aliases. In addition, the peak in this AoV spectrum is much narrower and, therefore, provides much higher period precision.

The periods determined from the AoV spectra are shown in Table~\ref{table4}. To find the precise peak frequency in each AoV spectrum, I used a Gauss function fit to the peak, which was cut off from the bottom by 40 \%.  Unfortunately, as seen in Fig.~\ref{figure2}, the noise in the AoV spectra is distorted by aliases over a wide frequency range. So I determined the average noise level in each AoV spectrum by averaging only 15 \% of the AoV spectrum at the beginning and 15 \% of the AoV spectrum at the end. Then, using a Gauss function fit to the peak, I determined its half-width at the $S-N$ level, where $S$ is the peak height and $N$ is the average noise level, and accepted it as the rms error of the period in accordance with the method by \citet{schwarzenberg91}. I also measured the half-width of the peak at half-maximum (HWHM). The HWHM of the peaks were about ten times greater than the errors determined using the method by \citeauthor{schwarzenberg91}. The most precise orbital period determined from the AoV spectrum of all the data combined is 0.188\,2118(8) d. To measure the peak frequency and its half-width, I also used a Lorentz function fit to the peak and obtained similar results.

\begin{table}[t]
\caption[ ]{Determination of the period from pairs of eclipses separated by large time intervals }
\scriptsize
\label{table5}
\begin{tabular}{@{}l c c c}
\noalign{\smallskip}
\hline
\noalign{\smallskip}
\tiny BJD$_{\rm TDB}$ mid-ecl.\,1 &\tiny BJD$_{\rm TDB}$ mid-ecl.\,2 &  Number   & Period (d) \\
 (-2,459,000)                                 &   (-2,459,000)                                &  of cycles  &                 \\
\noalign{\smallskip}
\hline
\noalign{\smallskip}
103.35356(18)     &    311.32775(33)    &   1105    & 0.18821194(34)     \\
111.25910(50)     &    322.24379(18)    &   1121    & 0.18821114(47)     \\
117.28177(27)      &   321.30262(31)    &   1084    & 0.18821112(38)     \\
117.47005(37)      &   322.43244(21)    &   1089    & 0.18821156(39)     \\
118.22241(47)     &    323.37294(13)    &   1090    & 0.18821150(45)     \\
\hline
\noalign{\smallskip}
                                  & & Average    &                                  \\
                                  & & period :      &    0.18821145(15)   \\
\noalign{\smallskip}
\hline
\end{tabular}
\end{table}

\begin{table}[b]
\caption[ ]{Verification of the ephemeris}
\small
\label{table6}
\begin{tabular}{@{}l c c c}
\noalign{\smallskip}
\hline
\noalign{\smallskip}
Date (UT) & \tiny BJD$_{\rm TDB}$ mid-ecl.  & Number          &  O--C$\times 10^{3}$ (d) \\
                  & (-2,459,000)                               & of cycles         &                 \\
\noalign{\smallskip}
\hline
\noalign{\smallskip}
2020 Sep. 10    & 103.35356      &      0          & $0.00\pm0.18$      \\
2020 Sep. 18    & 111.25910      &      42        & $0.66\pm0.50$      \\
2020 Sep. 24a  & 117.28177      &      74        & $0.56\pm0.27$      \\
2020 Sep. 24b  & 117.47005      &      75        & $0.63\pm0.37$      \\
2020 Sep. 25a  & 118.22241      &      79        & $0.15\pm0.47$      \\
2020 Sep. 25b  & 118.41034      &      80        & $-0.14\pm0.55$     \\
2021 Apr. 6       & 311.32775      &     1105     & $0.54\pm0.33$      \\
2021 Apr. 16     & 321.30262      &     1158     & $0.20\pm0.31$      \\
2021 Apr. 17a   & 322.24379      &     1163     & $0.31\pm0.18$      \\
2021 Apr. 17b   & 322.43244      &     1164     & $0.75\pm0.21$      \\
2021 Apr. 18    & 323.37294       &     1169      & $0.19\pm0.13$     \\
\noalign{\smallskip}
\hline
\end{tabular}
\end{table}

The precision of the period determined from the AoV spectrum of all the data combined is sufficient to obtain a tentative eclipse ephemeris, which can be used to fit to the eclipse times.  Indeed, if I consider the period error determined using the method by \citet{schwarzenberg91}, the accumulated error over the 7 months of my observations will reach 0.001 of the orbital phase. If I consider the period error equal to HWHM, which is obviously overestimated, the accumulated error over the 7 months of my observations will reach 0.01 of the orbital phase. Nevertheless, I additionally used independent pairs of eclipses separated by large time intervals to increase the precision of the period. To equalize the precision obtained from different pairs, I combined high-quality and low-quality eclipses. The results are shown in Table~\ref{table5}. The average orbital period obtained from 5 independent pairs of eclipses is 0.188\,211\,45(15)~d. 

\subsection{Eclipse ephemeris}

Using the mid-eclipse time of the first high-quality eclipse and the orbital period determined from 5 pairs of eclipses separated by large time intervals, I obtained the following tentative ephemeris:

{\scriptsize
\begin{equation}
\small BJD_{\rm TDB}(\rm mid-ecl.)=245\,9103.353\,56(18)+0.188\,211\,45(15) {\it E}. 
\label{ephemeris1}
\end{equation} }

\begin{figure}[t]
\includegraphics[width=84mm]{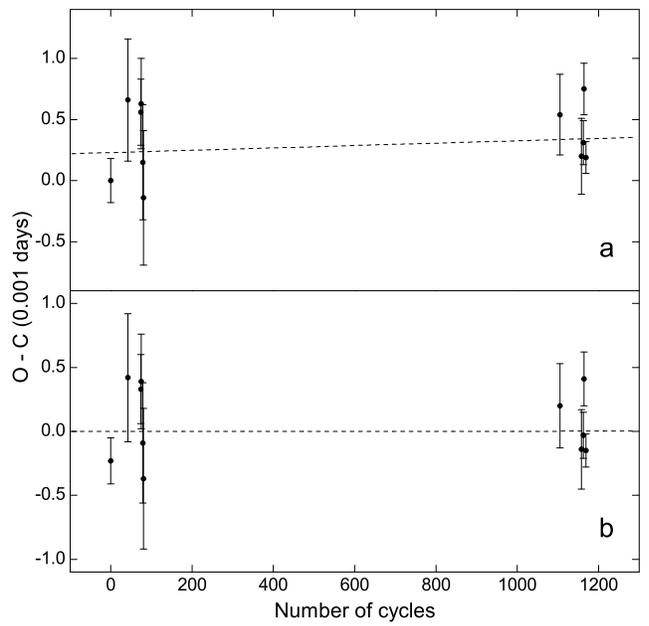}
\caption{(a) (O--C) diagram calculated for the tentative ephemeris obtained using the time of the first eclipse and the orbital period determined from pairs of eclipses separated by large time intervals. This (O--C) diagram shows the slope and displacement along the ordinate. (b) (O--C) diagram calculated for the ephemeris, which is corrected using the linear fit of the tentative ephemeris
}
\label{figure3}
\end{figure}

Using ephemeris~\ref{ephemeris1}, I calculated $O-C$, which are presented in Table~\ref{table6}. Their errors include only the errors of the observed time and do not include the errors of the calculated time determined by the errors of the ephemeris coefficients because the true errors of the ephemeris coefficients should be obtained from fitting. Table~\ref{table6} contains all the eclipses of high, moderate and low quality. This is evident from the errors of $O-C$, which differ noticeably from each other. So I performed a linear fit of $O-C$ using the errors as weights. The (O--C) diagram is presented in Fig.~\ref{figure3}(a). It shows the noticeable slope and displacement along the ordinate. $O-C$ obey the linear relation: $O - C = 0.000\,23(13) + 0.000\,000\,10(14) {\it E}$. Although the coefficients of this relation are comparable with their errors, I yet corrected ephemeris~\ref{ephemeris1} and obtained the final ephemeris:

{\scriptsize
\begin{equation}
\small BJD_{\rm TDB}(\rm mid-ecl.)=245\,9103.353\,79(13)+0.188\,211\,55(14) {\it E}. 
\label{ephemeris2}
\end{equation} }



\begin{table}[t]
{\scriptsize 
\caption{The periods obtained from different methods}
\label{table7}
\begin{tabular}{@{}l l c}
\noalign{\smallskip}
\hline
\noalign{\smallskip}
Method & Period (d) & Deviation \\           
                              & & from linear fit           \\
\noalign{\smallskip}
\hline
\noalign{\smallskip}
3 pairs of consecutive eclipses           & 0.18829(21)           & 0.4$\sigma$ \\ 
AoV spectrum of all the data             & 0.1882118(8)         & 0.3$\sigma$ \\
5 Pairs of separated eclipses           & 0.18821145(15)     & 0.5$\sigma$ \\ 
Linear fit of the ephemeris                & 0.18821155(14)     & -- \\  
\noalign{\smallskip}
\hline
\noalign{\smallskip}
\end{tabular}}
\end{table}

\begin{figure}[t]
\includegraphics[width=84mm]{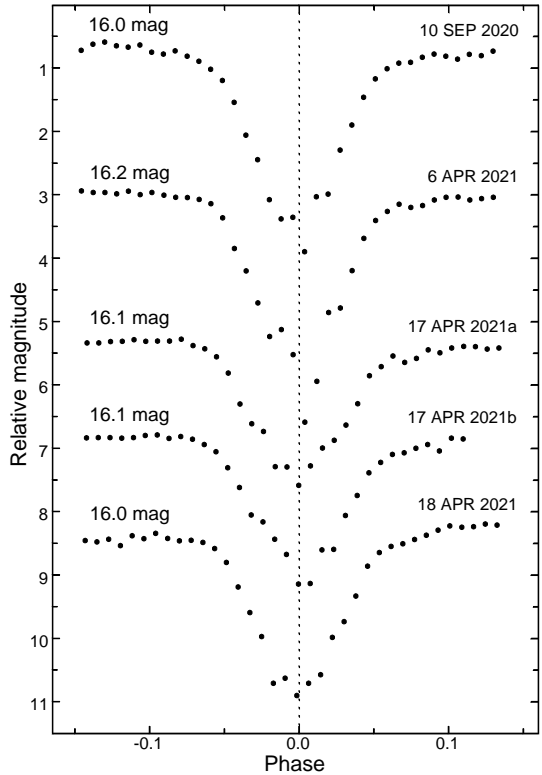}
\caption{Detailed view of 5 high-quality eclipses that were obtained in the high-brightness states of ZTF17 (16.0--16.2~mag). The eclipses are shifted for clarity. For each case, the out-of-eclipse magnitude is shown on the left. The orbital phases are calculated according to ephemeris~\ref{ephemeris2} }
\label{figure4}
\end{figure}

The (O--C) diagram obtained using ephemeris~\ref{ephemeris2} is shown in Fig.~\ref{figure3}(b). As seen, the slope and displacement along the ordinate are completely excluded. The orbital period obtained from the linear fit of ephemeris 1 to the mid-eclipse times is 0.188\,211\,55(14)~d. This period and its rms error are close to the orbital period and to its rms error obtained from the 5 pairs of eclipses separated by large time intervals. This does not seem surprising because in both cases nearly the same number of eclipses was used and these eclipses covered nearly the same time span. 

Table~\ref{table7} shows the periods determined using various methods and their deviations from the period determined from the linear fit of the ephemeris. Within the rms errors, all periods are compatible with each other. However, the period error determined from the AoV spectrum of all the data using the method by \citet{schwarzenberg91} is 6 times greater than the period error determined from the linear fit of the ephemeris. This suggests that the method by \citeauthor{schwarzenberg91} applied to the AoV spectrum gives an overestimated error. Indeed, if I assume that the true error of the period determined from the AoV spectrum is the same as the period error determined from the linear fit, then the deviation of this period from the period determined from the linear fit will be $1.3\sigma$. This is an admissible deviation.

The time during which the accumulated error from the period reaches one oscillation cycle is considered the validity time of the ephemeris. This time means the time during which the ephemeris can be used without ambiguity in determining the number of cycles. According to the rms error of the orbital period, the validity time of ephemeris~\ref{ephemeris2} is 700 years (a $1\sigma$ confidence level).

Fig.~\ref{figure4} shows the detailed view of 5 high-quality eclipses arranged according to the orbital phase calculated using ephemeris~\ref{ephemeris2}. As seen, the eclipses do not show phase deviations. The depth of these eclipses varies noticeably in the range 2.1--2.9~mag (see Table~\ref{table2}). These variations do not depend on the brightness of ZTF17 outside the eclipses, which was in the narrow range 16.0--16.2~mag.

\subsection{Folded light curves}

\begin{figure*}[t]
\includegraphics[width=174mm]{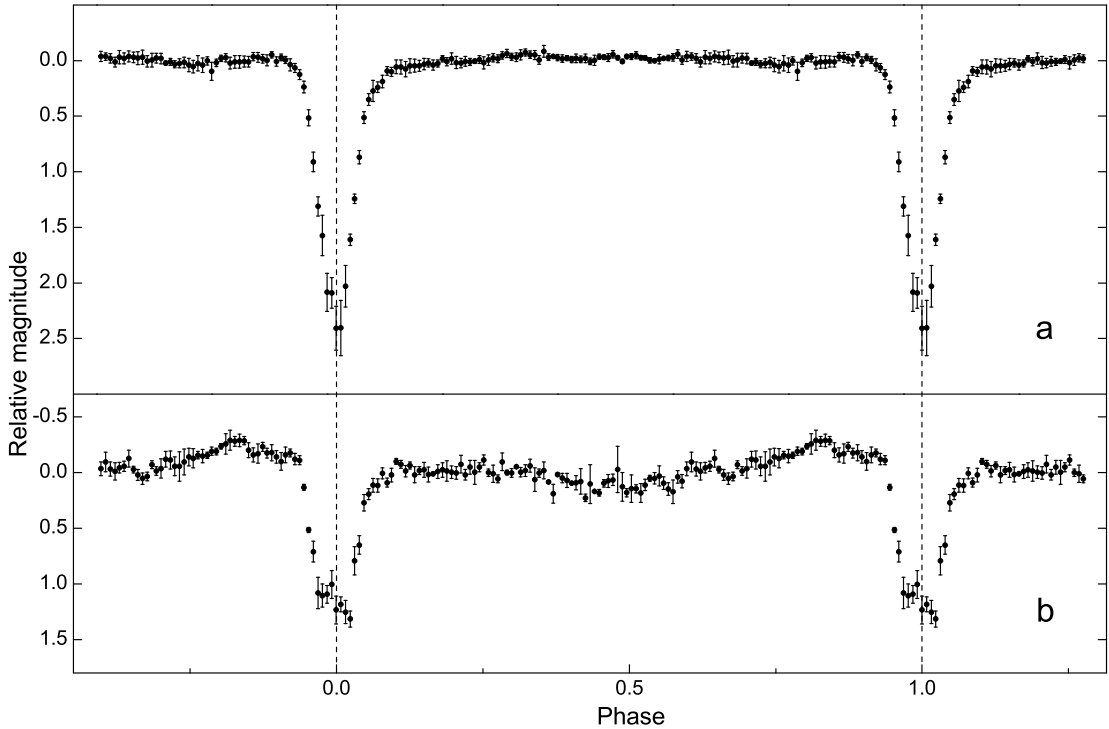}
\caption{Light curves folded with a period of 0.18821155~d.  (a) Folded light curve containing 5 eclipses obtained in the high-brightness states of ZTF17 (16.0--16.2~mag) and 1 eclipse obtained in the moderate-brightness state of ZTF17 (16.9~mag). (b) Folded light curve containing 5 eclipses obtained in the low-brightness state of ZTF17 (17.2--17.4~mag). The orbital phases are calculated according to ephemeris~\ref{ephemeris2} }
\label{figure5}
\end{figure*}

Fig.~\ref{figure5} shows the light curves of ZTF17 folded with the orbital period. The data were grouped into 127 phase bins. So the time interval between the consecutive points in the folded light curves (128.04 s) nearly coincides with the time resolution. Fig.~\ref{figure5}(a) shows the folded light curve obtained in the high- and moderate-brightness states of ZTF17 (the average out-of-eclipse magnitude is $16.23\pm0.17$~mag, see Table~\ref{table2}). This folded light curve contains 5 high-quality eclipses and 1 moderate-quality eclipse. I included the moderate-quality eclipse in this group of eclipses because its profile was similar to the profiles of high-quality eclipses. Fig.~\ref{figure5}(b) shows the folded light curve obtained in the low-brightness state of  ZTF17 (the average out-of-eclipse magnitude is $17.27\pm0.05$~mag). This folded light curve contains 5 low-quality eclipses. As seen, the average profiles and depths of eclipses differ significantly in these two cases. In case (a), the eclipse has a V-shape and a depth of $2.35\pm0.03$~mag. In case (b), the eclipse has a U-shape and a depth of $1.45\pm0.06$~mag. The depths were determined using a Gauss function fit to folded light curves.

As seen in Fig.~\ref{figure5}, the averaged eclipses show asymmetric wings. The careful examination of the eclipse profiles using a Gauss function fit proves that this asymmetry is statistically significant. Indeed, in Fig.~\ref{figure5}(a), the left eclipse wing has 3 consecutive points that are above the Gauss curve by 1$\sigma$, and the right eclipse wing has 5 consecutive points that are below the Gauss curve by 1--2$\sigma$. In Fig.~\ref{figure5}(b), the left eclipse wing has 5 consecutive points that are above the Gauss curve by 2--3$\sigma$, and the right eclipse wing has 4 consecutive points that are below the Gauss curve by 1--3$\sigma$.  

Because of the noise, it is difficult to measure the accurate duration of eclipses including asymmetric wings. However, using a Gauss function fit, I measured the widths of the prominent parts of the averaged eclipses at the level that is 0.3~mag lower than the brightness outside the eclipses. At this level, the asymmetric eclipse wings are imperceptible. The widths of the prominent parts of the averaged eclipses are $30.3\pm0.5$ and $32.9\pm1.7$~min. in Fig.~\ref{figure5}(a) and in Fig.~\ref{figure5}(b), respectively. Their difference is $2.6\pm1.8$~min. and is not statistically significant. Thus, the prominent parts of the averaged eclipses in the high- and low-brightness states of ZTF17 have approximately the same width.

The errors of the innermost points of the averaged eclipse in the high- and moderate-brightness states of ZTF17, which are visible in Fig.~\ref{figure5}(a), are noticeably greater compared to the errors of the innermost points of the averaged eclipse in the low-brightness state of ZTF17, which are visible in Fig.~\ref{figure5}(b). In addition, the eclipse depth determined by averaging the depths of individual eclipses in the high-brightness states of ZTF17 (16.0--16.2~mag) is $2.50\pm0.18$~mag, and the eclipse depth determined by averaging the depths of individual eclipses in the low-brightness state of ZTF17 (17.2--17.4~mag) is $1.45\pm0.06$~mag (see Table~\ref{table2}). Here, I excluded the moderate-quality eclipse. In this case, the depth error in the high-brightness states of ZTF17 is also noticeably greater. These greater errors indicate that the depth changes of the high-quality eclipses that are visible in Fig.~\ref{figure4} are real and cannot be caused by inaccurate subtraction of the sky background. Indeed, these errors should be approximately the same because in both cases, i.e. when ZTF17 was bright, and when ZTF17 was faint, the average magnitudes of the innermost points of the eclipses were nearly the same, $18.6\pm0.2$ and $18.7\pm0.1$~mag, respectively (see Table~\ref{table2}).


\subsection{Rapid variability}

I analysed the light curves obtained with a time resolution of 16~s to search for rapid periodic oscillations similar to those in intermediate polars. I used 5 light curves obtained when ZTF17 was brighter than 17~mag because these light curves had lower photon noise. Eclipses were excluded from these light curves, and the corresponding gaps were filled with points using linear interpolation between the magnitudes before and after the eclipses. The nightly averages and slopes in these light curves were removed using a linear fit. The total duration of these light curves was 20 h. I calculated the power spectrum for each individual light curve using the fast Fourier transform algorithm.

The averaged power spectrum obtained by averaging 5 individual power spectra clearly showed the second, third, fourth and fifth harmonics of the orbital period. At higher frequencies from 0.46~mHz up to Nyquist frequency (periods from 36~min. to 32~s), this power spectrum showed only white noise. The maximum peak with a semi-amplitude of 15~mmag appeared with a period of 360~s. This peak only slightly exceeded a confidence level of 99.95 \%. This is not sufficient to consider it statistically significant. Moreover, this peak was only 10 \% higher than other noticeable peaks that randomly appeared at frequencies above 0.46 mHz and were obviously caused by noise. Thus, in the period range from 36~min. to 32~s, ZTF17 did not show periodic oscillations with semi-amplitudes exceeding 15~mmag.

The averaged power spectrum described above did not show red noise that could indicate flickering. However, the red noise can be masked by a large number of harmonics of the orbital period. The out-of-eclipse light curves obtained in the high-brightness states of ZTF17 are quite smooth and do not show obvious rapid changes that could be attributed to flickering (e.g., the lower frame of Fig.~\ref{figure1}). The out-of-eclipse light curves obtained in the low-brightness state (e.g., the upper and middle frames of Fig.~\ref{figure1}) seem to hint at such rapid changes. However, these light curves have greater photon noise. Therefore, flickering in these light curves should be less noticeable.  To objectively confirm the flickering in ZTF17, I analyzed the out-of eclipse light curves to find out if they contain flickering-induced noise in addition to photon noise. During the observations, all counts of the two stars and the sky background are stored in the computer's hard disk. Therefore, the photon noise is easy to calculate. In addition, because ZTF17 is a faint star, other noise components, such as the noise caused by atmospheric scintillations and the noise caused by the movement of stellar images in the photometer diaphragms, should be relatively low (see, e.g., \citealt{kozhevnikoviz}).

I analyzed the light curves of ZTF17 obtained with a time resolution of 128~s. From each night of observations, I separated the largest continuous segment of the light curve without an eclipse and removed changes in the orbital time scale by subtracting a third-order polynomial fit. In total, there were 8 segments of the light curves lasting 1.6--3.9~h. For each segment, I calculated the standard deviation of the points in the light curve, the photon noise and their ratio. For 4 segments of the light curves obtained in the high-brightness states of ZTF17 (16.0--16.2 mag), the average photon noise was $0.027\pm0.002$~mag and the average ratio of the standard deviation of the points to the photon noise was $1.46\pm0.12$. The deviation from 1 is statistically significant because it reaches 3.8$\sigma$. Taking into account that the total dispersion of the points of the light curve is the sum of the squares of the various noise components (e.g., \citealt{kozhevnikoviz}), I determined that the flickering-induced noise (rms) is $0.029\pm0.003$~mag. Thus, noise analysis detects flickering in the high-brightness states of ZTF17, where the flickering-induced noise is nearly equal to the photon noise. For 4 other segments of the light curves obtained in the moderate- and low-brightness states of ZTF17 (16.9--17.4~mag), the average photon noise was $0.056\pm0.003$~mag and the average ratio of the standard deviation of the points to the photon noise was $1.67\pm0.34$. The deviation from 1 is not statistically significant because this deviation only reaches 2.0$\sigma$. Thus, noise analysis does not allow me to detect flickering in the moderate- and low-brightness stats of ZTF17. The possible reasons may be unstable flickering amplitude and large photon noise.

\begin{figure*}[t]
\includegraphics[width=174mm]{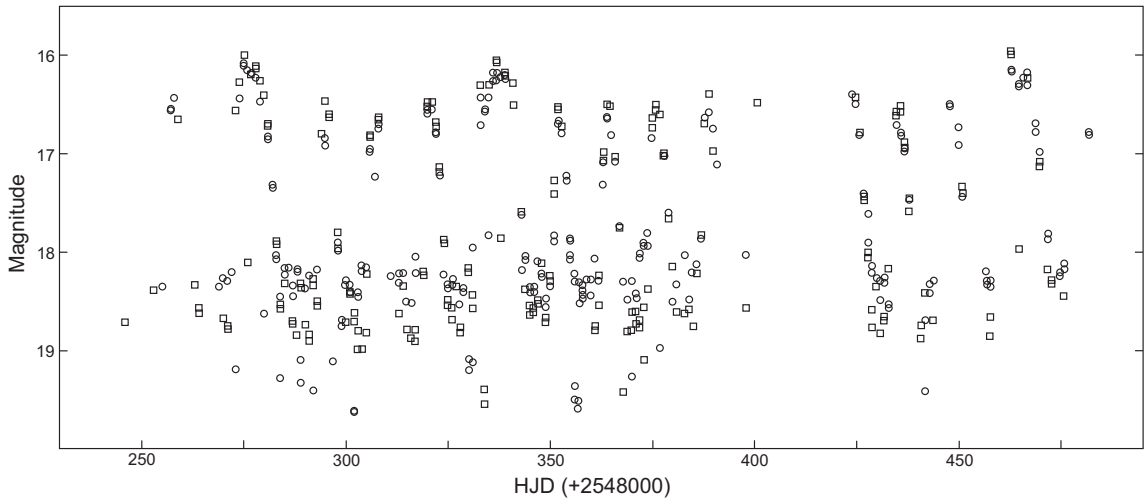}
\caption{Long-term light curve of ZTF17 obtained using the Zwicky Transient Facility survey. The magnitudes in the g-band are shown as squares. The magnitudes in the r-band are shown as circles. The magnitudes in the r-band are shifted by 0.5~mag to the g standard.}
\label{figure6}
\end{figure*}

I observed the intermediate polar V2069~Cyg using the same technique \citep{kozhevnikov17}. Although V2069~Cyg is only 0.5~mag brighter than ZTF17, its light curves clearly show flickering (see Fig.~1 in \citealt{kozhevnikov17}). To find out the conditions that allow me to see the flickering directly in the light curves, I analyzed the noise of the data of V2069~Cyg in the same way as I did for ZTF17. I used two long V2069~Cyg light curves obtained in 2014 November 16 and in 2015 September 13 (see Fig.~1 in \citealt{kozhevnikov17}). The points of these light curves were averaged over 128~s time intervals. From each light curve, I separated 2 segments lasting 2.9--3.4~h and removed possible changes in the orbital time scale by subtracting a third-order polynomial fit. For each segment, I calculated the standard deviation of the points in the light curve, the photon noise and their ratio. In addition to the raw light curves, I used the corresponding light curves from which the spin oscillation was subtracted. I did not find any noticeable difference. For the two light curves of V2069~Cyg, the photon noise was the same, 0.014~mag. The average ratio of the standard deviation of the points to the photon noise was $4.9\pm0.2$. From the photon noise and the standard deviation of the points, I determined that the flickering-induced noise is $0.067\pm0.003$~mag. Thus, in V 2069~Cyg, the flickering-induced noise is 4.8 times greater than the photon noise. Therefore the flickering is easily visible in the light curves.  

As mentioned, in the high-brightness states of ZTF17, the flickering-induced noise is nearly equal to the photon noise. If I assume that the flickering amplitude in ZTF17 is the same as in V2069~Cyg, the flickering-induced noise will be 2.5 times greater than the photon noise. This seems to be enough to directly distinguish the flickering in the light curves. However, in the moderate- and low-brightness states of ZTF17, the flickering-induced noise will be only 1.2 times greater than the photon noise, and the flickering will again be invisible in the light curves. Thus, in ZTF17, the photon noise is relatively high and the flickering amplitude is relatively low, and therefore the flickering itself is directly invisible. For stars fainter than 18~mag, the photon noise can be 0.2--0.4~mag (128~s time resolution). Even if the flickering amplitude in such faint stars is much larger than the flickering amplitude in V2069~Cyg, the photon noise will still be greater than the flickering-induced noise. Therefore, the flickering will be completely hidden.

\section{Discussion}

I performed photometric observations of the CV candidate ZTF17 and discovered very deep eclipses that were not previously known. In total, I observed 11 eclipses. The eclipses show asymmetric wings. Such asymmetric wings seem to be typical of eclipses observed in CVs. I observed such asymmetric eclipse wings in two CVs (see Fig. 4 in \citealt{kozhevnikov20} and Fig. 4 in \citealt{kozhevnikov18}). These asymmetric eclipse wings may be caused by the asymmetry of the accretion disc. Obviously, the asymmetric wings cannot be typical of eclipses observed in an ordinary eclipsing binary consisting only of stars. Based on the shape of the eclipses, I come to the conclusion that this CV candidate is a real CV. Taking into account the 6 outbursts detected by \citet{szkody20} in ZTF17 over 250 days, I conclude that this CV belongs to the dwarf nova subtype. 

\subsection{Outburst amplitudes}

The states of high and low brightness that I observed in ZTF17 differ only by 1.4~mag. This difference is too small and does not agree with the outburst amplitudes observed in dwarf novae. I probably observed ZTF17 in brightness states that do not strictly coincided with the outburst maximum or the quiescent state.  Fortunately, the International Variable Star Index (VSX) database contains a link to the long-term light curve of ZTF17 https://www.aavso.org/vsx/index.php?view=detail. \\ top\&oid=702847), which was obtained by Mariusz Bayer using the Zwicky Transient Facility survey (https://www.ztf.caltech.edu). This light curve shown in Fig.~\ref{figure6}  can solve the problem with the outburst amplitudes in ZTF17. It clearly shows 14 outbursts, each of which lasts 5--10 days, and the outburst amplitudes in the g-band are 2.1--2.6~mag. 

The total time interval of observations visible in Fig.~\ref{figure6} is approximately 190 days. Dividing this time interval by the number of outbursts, I determined that the outburst interval is 13.6 days. Using the Kukarkin-Parenago relation for dwarf novae (equation 3.1 in \citealt{warner95}), I found out that the outburst amplitudes in ZTF17 should be $2.8\pm0.7$mag. Thus, the outburst amplitudes visible in the long-term light curve obtained by Mariusz Bayer (Fig.~\ref{figure6}) are consistent with the Kukarkin-Parenago relation. In addition, after studying Fig. 3.8 in \citeauthor{warner95}, I came to the conclusion that ZTF17 has a very short outburst interval compared to other dwarf novae.

\subsection{Eclipse depths and profiles}

As seen in Fig.~\ref{figure4}, the eclipse depth varied noticeably in the range 2.1--2.9~mag when ZTF17 was bright (see Table~\ref{table2}). These variations occurred independently on the brightness of ZTF17 outside the eclipses, which was in the narrow range 16.0--16.2~mag. I observed similar variations in the eclipse depth in IPHAS J051814.33+294113.0 and J013031.89+622132.3, which did not correlate with small changes in the brightness of these CVs in their high-brightness states (see Fig.~5 in \citealt{kozhevnikov18} and Table~2 in \citealt{kozhevnikov20}. Therefore, noticeable variations in the eclipse depth in the high-brightness states of ZTF17 do not seem unusual. Such variations in the eclipse depth can be explained by changes in the distribution of brightness across the disc or changes in its radius, provided that the overall brightness of the disc shows only minor changes.

Whereas noticeable variations in the eclipse depth (Fig.~\ref{figure4}) occurred independently on the brightness when ZTF17 was bright (16.0--16.2~mag), the strong decrease in the average eclipse depth from $2.35\pm0.03$~mag to $1.45\pm0.06$ was evidently associated with the strong decrease in the brightness of ZTF17 to 17.2--17.4~mag because the eclipse profile also changed significantly from a V-shaped profile to a U-shaped profile. This is visible in Fig.~\ref{figure5}. Similar changes in the depth and profile of eclipses in EX~Dra were observed by \citet{baptista00} during the outburst cycle. Moreover, the U-shaped eclipse profile with a nearly flat bottom indicated that the eclipses in the quiescent state of EX~Dra were close to total eclipses. The width of the prominent parts of the eclipses in EX~Dra noticeably decreased in the quiescent state compared to the outburst, indicating that the radius of the disc also decreased (see Fig. 3 in \citealt{baptista00}).  This obviously led to a U-shaped eclipse profile. In contrast, the width of the prominent parts of the eclipses in ZTF17 did not noticeably change with the significant changes in the brightness of ZTF17 (Fig.~\ref{figure5}). Therefore, despite the fact that the eclipses observed in the low-brightness state of ZTF17 show a roughly flat bottom, which is visible in Fig.~\ref{figure5}(b), it seems premature to assert that these eclipses are close to total eclipses.

Another reason that can lead to flat-bottomed eclipses may be the effect of faint stars around ZTF17. Looking at PanSTARRS maps in the Aladin sky atlas (https://aladin.u-strasbg.fr/AladinLite), I found out three stars at a distance of 4.6--2.7 arcsec from ZTF17, which should be located inside the photometer diaphragm (16~arcsec). These three stars measured using the space observatory {\it Gaja} have $G$-magnitudes of 20.36, 20.54 and 20.56~mag. Although each of these three stars is very faint, their summary brightness is 19.29~mag and affect noticeably the innermost eclipse points. The average magnitude of the innermost eclipse points in the low-brightness state of ZTF17 is $18.7\pm0.1$~mag (see above). Having excluded the light of these three stars, I found that the corrected average magnitude of the innermost eclipse points should be in the range 19.3--19.9~mag. Diluting the light with these three stars can lead to flat-bottomed eclipses because the brightness change of ZTF17 in the range 19.3--19.9~mag will lead to a summary brightness change of only 0.26~mag. The eclipse depth in the high-brightness states of ZTF17 may also be greater than observed. The true eclipse depth and its profile can be determined using future CCD photometry in conditions of good seeing. Indeed, the optimal aperture for CCD photometry is 1--1.5~arcsec (see, e.g., Fig.~6 in \citealt{howell89}), and the effect of these three stars can be excluded.

The average eclipse depth in the high-brightness states of ZTF17 was $2.50\pm0.18$~mag. These eclipses seem to be very deep. To find out how often such deep eclipses occur among dwarf novae, I studied the International VSX database and references therein (https://www.aavso.org/vsx) as well as references in ADS (https://ui.adsabs.harvard.edu/classic-form).  I found 108 eclipsing dwarf novae with known eclipse depths. Only 15 of them show eclipses with depths noticeably exceeding 2~mag. In addition, only 3 of these 15 dwarf novae have orbital periods greater than the orbital period of ZTF17. These are AR~Cnc having $P_{\rm orb}=0.2146$~d and the eclipse depth of about 3~mag in quiescence \citep{howell90}, EX~Dra having $P_{\rm orb}=0.2099$~d and the eclipse depth reaching 2.4~mag in outburst (see Fig.~3 in \citealt{voloshina21}) and IPHAS~J051814.33+294113.0 having $P_{\rm orb}=0.2060$~d and the average eclipse depth of $2.42\pm0.06$~mag \citep{kozhevnikov18}. Recently, IPHAS~J051814.33+294113.0 was classified as a dwarf nova (http://ooruri.kusastro. kyoto-u.ac.jp/mailarchive/vsnet-chat/8502). Thus, the discovery of deep eclipses in ZTF17 is a valuable find because such deep eclipses are rare and because ZTF17 has a very long orbital period.

\subsection{Further studies}

The discovered eclipses allowed me to surely measure the orbital period, which is the most important parameter of any binary star, $P_{\rm orb}=0.188\,211\,55\pm0.000 \,000\,14$~d. The high precision of the period is achieved due to the observations of the large number of eclipses separated by large time intervals reaching 220 days. Although, as seen in Fig.~\ref{figure3}, two compact groups of observed eclipses are separated by the large gap in which no eclipses were observed, this does not cause problems with aliases (Fig.~\ref{figure2}). Obviously, sharp eclipses are favourable to avoid such problems. I derived the eclipse ephemeris, which, according to the precision of the orbital period, has a long validity time of 700 years. This ephemeris can be used for future studies of orbital period changes, which may be caused by Solar-like activity cycles of the red companion (e.g., \citealt{rubenstein91, applegate92}) or by a possible giant planet orbiting around the centre of gravity in ZFF17 (e.g., \citealt{bruch14, beuermann11}).

Deep eclipses in CVs make it possible to study the structure and evolution of the accretion disc over time using eclipse mapping methods \citep{horne85, baptista04}. This is especially interesting for ZTF17, which is a dwarf nova, because such studies allow us to track changes in the disc structure during the outburst cycle (e.g., \citealt{baptista01}). Although the presence of three faint stars around ZTF17 somewhat complicates the task of obtaining an accurate eclipse profile, nevertheless, such studies are possible using CCD photometry in conditions of good seeing. On the other hand, the study of the disc evolution during the outburst cycle is facilitated by the fact that outbursts in ZTF17 occur frequently. As the long-term light curve obtained by Mariusz Bayer shows (Fig.~\ref{figure6}), the outburst interval in ZTF17 is about 14 days.

Being an eclipsing binary, ZTF17 is suitable for determining the masses of its stellar components because the eclipses allow determining the inclination of the orbit. In addition, ZTF17 has a long orbital period, and, therefore, its parameters are particularly interesting.  Some dwarf novae show complex eclipse profiles with clear and separable white dwarf and bright-spot eclipses (see, e.g., Fig. 1 in \citealt{hardy17}). Then the physical model of the binary system can give reliable parameters of the system without any spectroscopic study (e.g., \citealt{littlefair14, wood86}).  However, as seen in Table 1 in \citet{zorotovic11}, this method is mainly used for systems with short orbital periods. Therefore, it seems unlikely that future CCD observations of ZTF17 will show such a complex eclipse profile.  Threfore, to determine the masses of the stellar components in ZTF17, radial velocity measurements should be used together with photometric data. Such measurements make it possible to simultaneously obtain stellar masses and the inclination of the orbit (e.g., \citealt{szkody93, downes86}). However, because the spectrum can be contaminated by asymmetric disc structures \citep{robinson92}, measurements of radial velocities are difficult. In future spectroscopic observations of ZTF17, the orbital phases determined from ephemeris~\ref{ephemeris2} can help to identify these structures and thus facilitate radial velocity measurements (e.g., Hellier, 2001).

\section{Conclusions}
I performed photometric observations of the CV candidate ZTF17 and discovered very deep eclipses in this star for the first time. A comprehensive analysis of my photometric data, which were obtained over 8 nights covering 7 months, gives the following results:

\begin{enumerate}

\item Due to the large observational coverage, I measured the orbital period in ZTF17 with high precision, $P_{\rm orb}=0.188\,211\,55\pm0.000\,000\,14$~d. 
\item I obtained the eclipse ephemeris.  It has a validity time of 700 years in accordance with the precision of the orbital period. This ephemeris can be used for future studies of orbital period changes in ZTF17.
\item During my observations, ZTF17 showed brightness changes between 16.0 and 17.4~mag resembling dwarf nova outbursts and quiescent states.
\item When ZTF17 was bright (16.0--16.2~mag), the eclipse depth varied in the range 2.1--2.9~mag, and the average eclipse depth was $2.50\pm0.18$~mag. When ZTF17 was faint (17.2--17.4~mag), the eclipse depth was less unstable, and the average eclipse depth was $1.45\pm0.06$~mag. However, these differences may be caused by the contaminating effect of three faint stars around ZTF17.
\item In both cases, i.e. when ZTF17 was bright, and when ZTF17 was faint, the average width of the prominent parts of the eclipses was approximately the same and was roughly 30~min.
\item Due to the long orbital period, ZTF17 is of interest for determining the masses of its stellar components. Because, in accordance with the long orbital period, the eclipse profile in ZTF17 should be featureless, the eclipse modelling cannot give reliable masses of the components.  Therefore, it is necessary to use radial velocity measurements to determine the stellar masses. Then my precise eclipse ephemeris can be useful for determining the orbital phases and thus can facilitate radial velocity measurements.

\end{enumerate}

\section{Data availability}

The datasets generated during and/or analysed during the current study are available from the corresponding author on reasonable request.

\section*{Acknowledgments}

This work was supported in part by the Ministry of Science and Higher Education of the Russian Federation, FEUZ-2020-0030, and by the Act no. 211 of the Government of the Russian Federation, agreement no. 02.A03.21.0006. This research has made use of NASA's Astrophysics Data System Bibliographic Services and the International Variable Star Index (VSX) database. The VSX database is operated at AAVSO, Cambridge, Massachusetts, USA. This research has made use of the VizeR catalogue access tool \citep{ochsenbein00} and the Aladin sky atlas developed at CDS, Strasbourg observatory, France \citep{bonnarel00, boch14}. This work has made use of data from the European Space Agency (ESA) mission
{\it Gaia} ({https://www.cosmos.esa.int/gaia}), processed by the {\it Gaia}
Data Processing and Analysis Consortium (DPAC,
{https://www.cosmos.esa.int/web/gaia/dpac/consortium}). Funding for the DPAC
has been provided by national institutions, in particular the institutions
participating in the {\it Gaia} Multilateral Agreement.

This version of the article has been accepted for publication, after peer review
but is not the Version of Record and does not reflect post-acceptance
improvements, or any corrections. The Version of Record is available online at:
http://dx.doi.org/10.1007/s10509-021-04015-4.
Use of this Accepted Version is subject to the publisher's Accepted Manuscript terms of use https://www. springernature.com/gp/open-research/policies/accepted-manuscript-terms.

\vspace{1.0cm}
Fig. 1  Three longest light curves of ZTF17, which show consecutive eclipses. Note that the out-of-eclipse magnitude varies greatly from night to night

\vspace{0.4cm}
Fig. 2  Analysis of Variance spectra of ZTF17 calculated for the data obtained in 2020 September, for the data obtained in 2021 April and for all the data combined. Note that the main peak in the AoV spectrum of all the data is much narrower

\vspace{0.4cm}
Fig. 3  (a) (O--C) diagram calculated for the tentative ephemeris obtained using the time of the first eclipse and the orbital period determined from pairs of eclipses separated by large time intervals. This (O--C) diagram shows the slope and displacement along the ordinate. (b) (O--C) diagram calculated for the ephemeris, which is corrected using the linear fit of the tentative ephemeris

\vspace{0.4cm}
Fig. 4  Detailed view of 5 high-quality eclipses that were obtained in the high-brightness states of ZTF17 (16.0--16.2~mag). The eclipses are shifted for clarity. For each case, the out-of-eclipse magnitude is shown on the left. The orbital phases are calculated according to ephemeris~\ref{ephemeris2}

\vspace{0.4cm}
Fig. 5  Light curves folded with a period of 0.18821155~d.  (a) Folded light curve containing 5 eclipses obtained in the high-brightness states of ZTF17 (16.0--16.2~mag) and 1 eclipse obtained in the moderate-brightness state of ZTF17 (16.9~mag). (b) Folded light curve containing 5 eclipses obtained in the low-brightness state of ZTF17 (17.2--17.4~mag). The orbital phases are calculated according to ephemeris~\ref{ephemeris2}

\vspace{0.4cm}
Fig. 6  Long-term light curve of ZTF17 obtained using the Zwicky Transient Facility survey. The magnitudes in the g-band are shown as squares. The magnitudes in the r-band are shown as circles. The magnitudes in the r-band are shifted by 0.5~mag to the g standard

\end{document}